\documentclass[a4paper,UKenglish,cleveref, autoref, thm-restate]{lipics-v2019}

\nolinenumbers
\usepackage{xcolor}
\usepackage{soul}
\sethlcolor{black}
\makeatletter
\newif\if@blind
\@blindtrue 
\if@blind \sethlcolor{black}\else
   
\fi

\title{ChatGPT is not a pocket calculator - \\
Problems of AI-chatbots for teaching Geography} 

\titlerunning{ChatGPT is not a pocket calculator} 

\author{
Simon Scheider
}{
Utrecht University, Department of Human Geography and Spatial Planning, The Netherlands 
}{
s.scheider@uu.nl
}{
https://orcid.org/0000-0002-2267-4810
}{}

\author{
Harm Bartholomeus
}{
Wageningen University, Laboratory of Geo-Information Science and Remote Sensing, The Netherlands
}{
harm.bartholomeus@wur.nl
}{
https://orcid.org/0000-0002-1905-7678
}{}

\author{
Judith Verstegen
}{
Utrecht University, Department of Human Geography and Spatial Planning, The Netherlands
}{
j.a.verstegen@uu.nl
}{
https://orcid.org/0000-0002-9082-4323
}{}

\authorrunning{
S. Scheider, H. Bartholomeus and J. Verstegen
} 

\Copyright{
Simon Scheider, Harm Bartholomeus and Judith Verstegen
} 



\acknowledgements{
We want to thank all our survey participants. The work was supported by the European Research Council (ERC) under the European Union’s Horizon 2020 research and innovation programme (grant agreement No 803498).
}

\hideLIPIcs  

\EventEditors{John Q. Open and Joan R. Access}
\EventNoEds{2}
\EventLongTitle{42nd Conference on Very Important Topics (CVIT 2016)}
\EventShortTitle{CVIT 2016}
\EventAcronym{CVIT}
\EventYear{2016}
\EventDate{December 24--27, 2016}
\EventLocation{Little Whinging, United Kingdom}
\EventLogo{}
\SeriesVolume{42}
\ArticleNo{23}

\begin{document}

\maketitle
\begin{abstract}
The recent success of large language models and AI chatbots such as ChatGPT in various knowledge domains has a severe impact on teaching and learning Geography and GIScience. The underlying revolution is often compared to the introduction of pocket calculators, suggesting analogous adaptations that prioritize higher-level skills over other learning content. However, using ChatGPT can be fraudulent because it threatens the validity of assessments. The success of such a strategy therefore rests on the assumption that lower-level learning goals are substitutable by AI, and supervision and assessments can be refocused on higher-level goals. Based on a preliminary survey on ChatGPT's quality in answering questions in Geography and GIScience, we demonstrate that this assumption might be fairly naive, and effective control in assessments and supervision is required. 
\end{abstract}

\section{Introduction}
\label{sec:intro}
The introduction of transformer-based large language models, such as GPT-3, and related chatbot services such as ChatGPT, has hit teachers hard on all levels of the educational system, from primary schools to universities. In the subsequent debate, voices in academia and practice range from alarm calls to appeals to remain calm and accept the unavoidable. In the latter camp, the argument goes that teachers should embrace this development and its supposed benefits for teaching\footnote{\url{https://www.computingatschool.org.uk/resource-library/2023/january/chatgpt-for-teachers-a-guide-by-evan-dunne}} rather than fight against the threats. If allowing ChatGPT in assignments, so the story goes, students and teachers can benefit from collaborative learning, personalized teaching material and a reduction of the workload for teachers \cite{dijkstra2022, kasneci2023chatgpt}. 

Sam Altman, the CEO of ChatGPT’s maker Open AI, used similar arguments when comparing the current situation to the introduction of pocket calculators\footnote{\url{https://www.yahoo.com/news/ceo-chatgpt-maker-responds-schools-174705479.html}}: 
\begin{quote}\small
Generative text is something we all need to adapt to [...] We adapted to calculators and changed what we tested for in math class, I imagine. This is a more extreme version of that, no doubt, but also the benefits of it are more extreme, as well.
\end{quote}

However, though there is no doubt that benefits exist, we believe the threats should be carefully considered before assuming that the former outweigh the latter, or that adaptation is the right strategy. In particular, general benefits of language models and related chatbot services for \textit{society as a whole} do not automatically translate to benefits for \textit{teaching}. The role of a teacher is to 1) teach students knowledge and skills, and, for this purpose, to 2) assess to what extent this has succeeded. If chatbot services can mimic these skills, then what is at stake in education is nothing less than the \textit{validity of assessments}, i.e., the way teachers make sure students have gained essential knowledge and skills. 

The severeness of the threat, therefore, \textit{is a function of the quality of answers} of the chatbot to questions in essential areas of knowledge and levels of skills. For this reason, we argue that finding out the real magnitude of the problem for education requires investigating the quality of ChatGPT for specific areas of knowledge and skills first. In this paper, we report on a preliminary assessment on ChatGPT's quality in \textit{teaching Geography and GIScience at university level}.

\section{The problem in a nutshell} \label{sec:problem}

\subsection{GPT-3 and ChatGPT}
The recent breakthrough in AI-based chatbots is largely due to the development of a particular generation of deep neural networks, called \textit{transformers} which are \textit{pretrained} on huge amounts of text. 
GPT-3 (Generative Pre-trained Transformer) is a transformer model that can produce texts of a human-like quality. \textit{Generative pre-trained} means that it has been trained in an unsupervised and unspecific way on vast amounts of texts to continue texts (words, code or data) given in some input prompt \cite{brown2020language}. GPT-3 was released in 2020 and was trained on 45 TB of text data from Wikipedia, book libraries and crawled web texts. The model has a size of roughly 800 GB and consists of 175 billion network parameters \cite{brown2020language}. Based on this model, the chatbot ChatGPT was fine-tuned in several stages, using human feedback as well as a reward model for reinforcement learning \cite{ouyang2022training}. The latter was mainly used to form an ''ethical shell'' to avoid that ChatGPT exhibits unethical behaviour \cite{broersen2023}.

As a result of these developments, the capacity of GPT-3 and ChatGPT to automatically answer questions, follow instructions, summarize and translate texts, correct grammar and generate new texts of excellent quality is unprecedented \cite{floridi2020gpt}. Furthermore, similar transformer models can not only be used to write prose and poetry, but also to automatically synthesize computer programs \cite{lu2021pretrained,brown2020language}, and for other modalities than text. 

\subsection{Is the use of ChatGPT in assessments fraud?}
Students can use AI chatbots to generate suggestions as input for their learning, and teachers can profit by generating learning or testing material\footnote{\url{https://theconversation.com/chatgpt-students-could-use-ai-to-cheat-but-its-a-chance-to-rethink-assessment-altogether-198019}}. So if ChatGPT is the next step after the pocket calculator revolution, eventually, it will become a fundamental cultural technique. 

This begs the question as to whether its use should be considered fraud. Academic dishonesty normally includes plagiarism, fabrication, falsification, misrepresentation, and misbehaviour \cite{burke2018applying}. ChatGPT does not clearly fall under any of these categories. Yet, it still leads to \textit{inaccurate measurements of skills}, thereby cheapening the degrees held by alumni, misleading employers, and diminishing moral integrity by being unfair to students who do not use it. To understand why, consider the following definition of fraud in the examination regulations of Utrecht University:
\begin{quote}\small
Fraud and plagiarism are defined as an action or omission on the part of the students which produces an incorrect representation of their performance as regards their knowledge, skills and understanding, which may result in an examiner no longer being able to assess the knowledge or ability of a student properly and fairly [...] (Paragraph 5 of the Education and Examination Regulations (EER) at Utrecht University, The Netherlands)
\end{quote}
While ChatGPT of course can play a constructive role in teaching\footnote{For example, in order to practice critical analysis of AI generated texts.}, \textit{its use in student assessments} inevitably falls under this definition, as soon as an examiner cannot distinguish anymore between student skills and ChatGPT skills, and thus \textit{cannot validly assess student skills according to  the learning goals}. 
This is independent of whether we assess learning goals in open- or closed-book exams or in home writing assignments \cite{mckeachie2013mckeachie}. 
It applies even when asking students to be creative, e.g., to criticize ChatGPT answers (as suggested in\footnote{\url{https://www.computingatschool.org.uk/resource-library/2023/january/chatgpt-for-teachers-a-guide-by-evan-dunne}}), since it can criticize its own answers, too \cite{ouyang2022training}. According to the \textit{fraud triangle} in academia \cite{burke2018applying}, students have an incentive for fraud due to work pressure, see opportunities for fraud when lacking institutional supervision, and tend to rationalize fraud a-posteriori by denying their responsibility. All this directly applies to ChatGPT: It releases students from work pressure (incentive),  is difficult to control due to lack of supervision, and fraud can be easily rationalized, in particular, by the reasons for questioning fraud given above.

\subsection{Why learning requires supervision}
If using AI  chatbots is fraudulent because it threatens the supervision of learners, then, one might argue, why not give up supervision in teaching? Yet, on closer inspection, this turns out to be a spurious argument. Despite the importance of autonomy in learning cognitive constructions \cite{richardson2003constructivist,glasersfeld.edu}, learning fundamentally requires guidance \cite{biesta2020risking}. Thus to acquire particular skills, students need instructions and explanations as well as feedback on the quality of their results \cite{mckeachie2013mckeachie}. Teachers have an essential role to play, not only by providing guidance but also by interrogating students\footnote{''Since the knowledge gained comes primarily through interrogation of and by others, education is \textit{relational}, depending on personal interaction between teacher and student.''\cite{gooch2019course}}. 

Because of this, the learning process cannot afford giving up on assessing essential skills in supervision. One can argue that some skills might lose their relevance when substituted by AI, while others become more important as cultural techniques. Yet, as a matter of fact, some skills always need to be learned, in particular, \textit{distinguishing valid from invalid answers}. And this is why the pocket calculator analogy breaks down: Mental arithmetic is a skill that is easily substitutable, yet choosing formulas to solve mathematical problems is not. 
Mental arithmetic is subordinate to mathematical problem solving since the latter is needed to judge whether an arithmetic formula can yield a valid answer in the first place. Thus, giving up on learning \textit{this} skill just means to give up on the very concept of validity \cite{janich2001logisch}. 
The question of substitutability boils down to whether the skills simulated by ChatGPT are (1) \textit{essential} or rather \textit{subordinate} to skills required by our culture to judge about a state of affairs, and (2) whether its \textit{quality is sufficient for substituting} these skills. To answer these questions, let us take a closer look at the concrete skills required in Geography and GIScience.


\section{ChatGPT quality in geographic exams and assignments}\label{sec:survey}
An OpenAI technical report \cite{openai2023} evaluated the performance of GPT-3.5 in publicly available US exams, mainly Advanced Placement (AP) tests; GPT-3.5 scored worst (0th-7th percentile of test-takers) in the AP Calculus BC and the best (91th-100th percentile of test-takers) in the AP Environmental Science. So, at least at college-level ChatGPT may be relatively good at answering geography questions. We have designed a survey among university teachers in Geography and GIScience to obtain their opinion on the quality of ChatGPT's answers to university exam and assignment questions.

\subsection{ChatGPT quality survey}

After an introduction to the study, an informed consent and sample-evaluation questions on age and gender, the participant was asked to select an exam question or assignment they would use for one of their courses and to pose this question/assignment to ChatGPT. We asked the participant to fill in the question and the answer, the name of the corresponding course, the educational level of the course (Bachelor 1, 2, 3, Master, PhD, or other). The answer of ChatGPT was then to be rated by the participant based on correctness, completeness, conciseness, and clarity, corresponding to question quality dimensions suggested in \cite{shah2010evaluating}. The survey was implemented in Qualtrics XM\footnote{\url{https://www.qualtrics.com/}}.

The participants were university teachers recruited across three departments at two universities: the Departments of Human Geography and Spatial Planning, and Physical Geography at Utrecht University, The Netherlands, and the Laboratory of Geoinformation Science and Remote Sensing at Wageningen University, The Netherlands. The survey was conducted between January 30th 2023, and April 21st 2023.

To analyse the results of the survey, we first filtered out invalid entries (e.g., missing or incomplete questions or answers). Next, we classified each question according to Bloom's (revised) taxonomy; from low to high level: remember, understand, apply, analyze, evaluate, create \cite{krathwohl2002}. This was done based on the verb in the question. Finally, we analyzed differences in quality scores (correctness, completeness, conciseness, and clarity) per geography domain, education level, and Bloom's taxonomy level.

\subsection{Survey results}

\begin{figure}[h]
\centering
\includegraphics[width=0.49\textwidth]{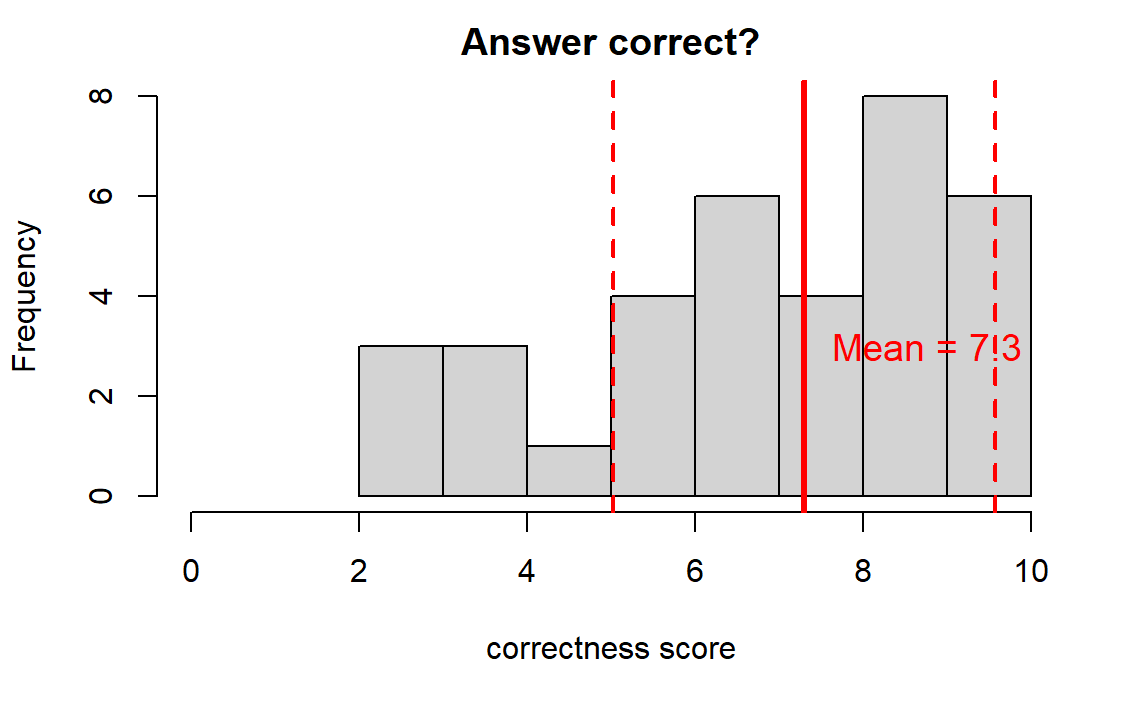}~
\includegraphics[width=0.49\textwidth]{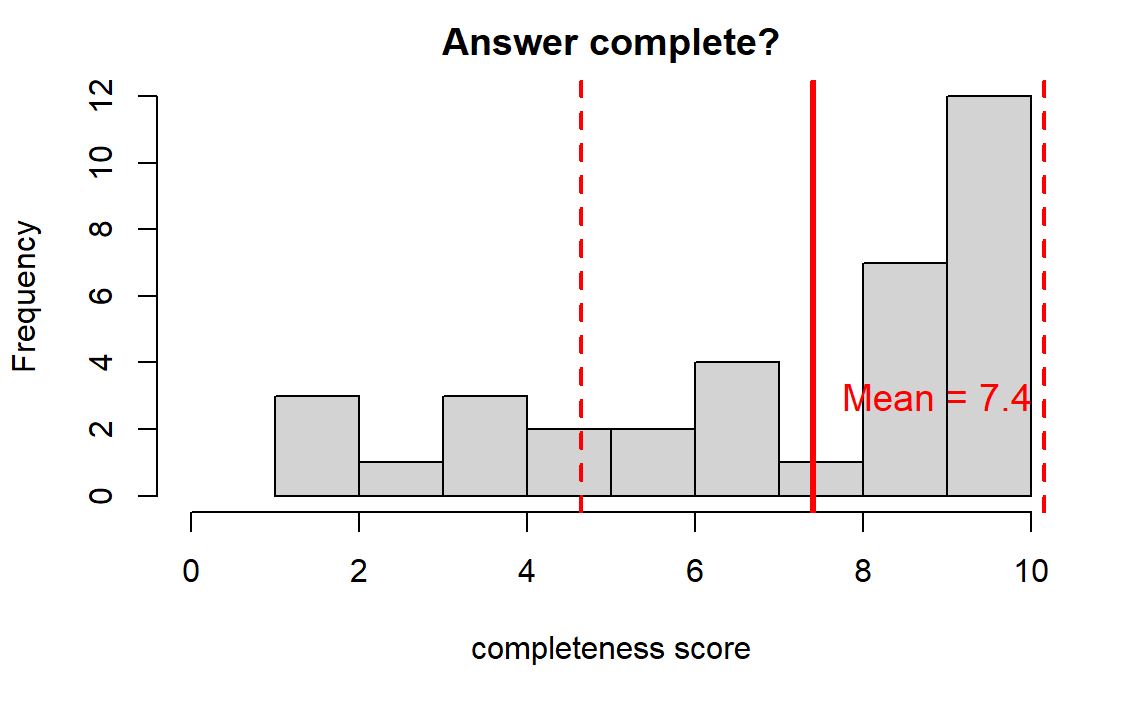}
\caption{Distribution of correctness and completeness scores for answers. Red lines show means and std deviations. }\label{fig:correctness}
\end{figure}

We had 41 valid survey responses, by 15 females, 26 males, and 0 other genders, with a median age of 45 years. Questions were distributed as following among the different knowledge domains (can be multiple per question): GIS and/or Remote Sensing (14), Urban Geography (9), Physical Geography (7), International Development Studies (6), Ecomonic Geography (3),  Hydrology (3), Spatial Planning (2) and Other (4). In the majority (80\%), the answers of ChatGPT have correctness scores that would allow it to pass the exam or assignment (score equal to or larger than 5.5 in Dutch exams), see Fig.~\ref{fig:correctness}. A bit less, $\approx$ 74\% of the answers are complete within the boundaries of the same threshold, and the same applies to clarity. The conciseness scores, in contrast, are much lower (47\%); many survey participants commented that the answers were overly long, beyond the point, and they would become tired if all students answered in this way. 

In contrast to what one may expect, the quality of the answers does not decrease with education level (BSc 1, 2, 3, MSc). No clear relation between education level of the question and quality of the answer appears from our survey. 

\begin{figure}[h]
\centering
\includegraphics[width=0.8\textwidth]{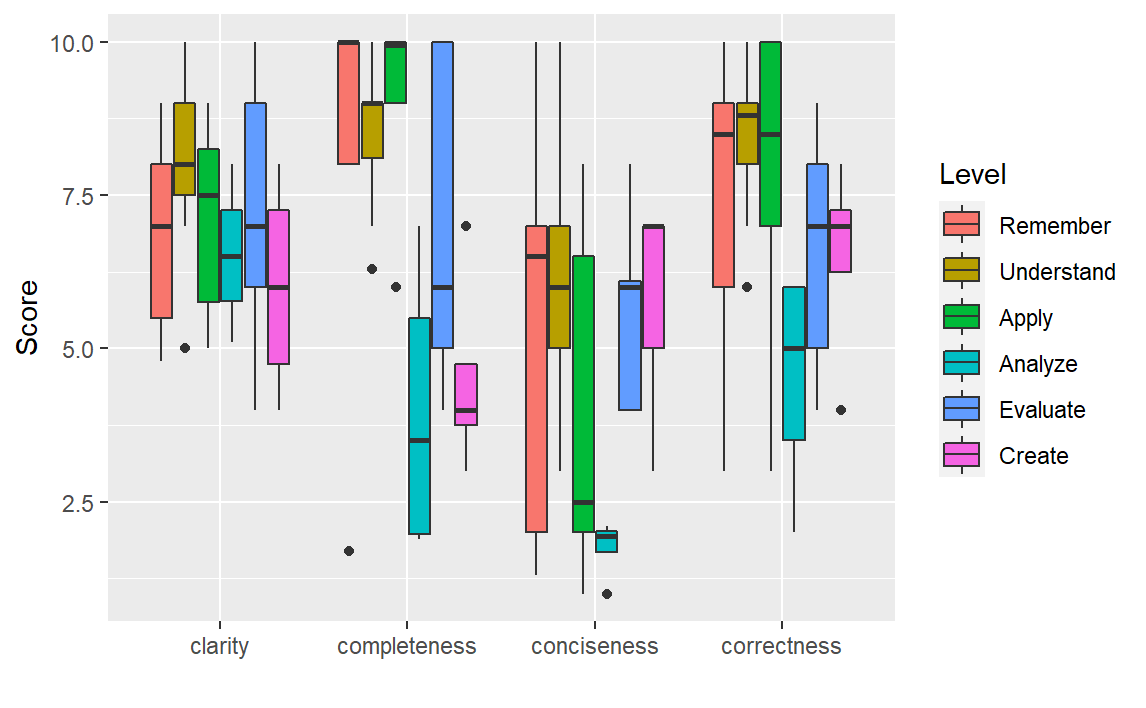}
\caption{Quality scores of answers to questions grouped by Bloom's taxonomy level.}\label{fig:bloom_level}
\end{figure}

In the levels of Bloom's taxonomy, a difference in quality appears between the lowest three and highest three Bloom skill levels (Fig.~\ref{fig:bloom_level}). A corresponding Wilcoxon rank sum test was significant for both correctness and completeness with \textit{p = .005}, having similar distribution shapes and sample sizes but a significant shift in the median between upper and lower levels. Questions that require students to \textit{evaluate}, \textit{analyze} and \textit{create} tend to be much less correct as well as complete. Yet, still, the correctness of corresponding answers tends to range above 5.5, and thus they may pass a test. Correspondingly, some survey participants commented that they were particularly worried about the capacity of ChatGPT to write essays and other types of assignments; this falls in the 'evaluate' or 'create' category. Assuming that the higher the skill level, the more essential the skills for learning, these results raise the serious concern that chatbots may pose a considerable threat to the validity of assessments in Geography exams and assignments.

\section{Discussion and conclusion}
Our preliminary results indicate that ChatGPT will likely pass Geography and GIScience exams and assignments in its present form; this is independent of the skill (Bloom's taxonomy) as well as the educational level. This means that even assessing only (true) academic skills does not help, as ChatGPT masters these too. In terms of correctness, the answers to Geography questions score satisfactorily (able to pass the exam) over different levels of education as well as different skill levels. The answers score slightly worse in terms of completeness and clarity, and much worse in terms of conciseness. Herein, note that we rated only the first answer, while ChatGPT can make an answer more concise if requested (but in our experience correctness then goes down). Higher Bloom levels are harder to match, but are on average still sufficiently correct for passing grades given the tested questions. Note that some questions at these levels could not be tested because they involve interpreting a provided graph or map, which cannot be pasted in the chat; therefore, our results may be biased towards higher grades. 

Since the skills underlying these higher levels are not substitutable when learning and thus require supervision, the introduction of ChatGPT is not comparable to the introduction of the pocket calculator. This means that we are forced to change Geography and GIScience assessments in one way or another. We can either:
\begin{enumerate}
    \item \textbf{Adapt the learning goals}, and thus the skills being assessed, following the way of thinking ''students do not need to learn the things that ChatGPT can do for them''. Since especially higher-level skills are essential for learning itself, we argue that this is not really an option for such skills. Be aware that the line of reasoning to ''embrace the development and allowing ChatGPT in assignments'' means, perhaps unintentionally, disregarding essential learning goals. What is and is not essential also depends on what society demands. 
    \item \textbf{Control the assessment environment}, for example by preventing access to the internet during the assessment. This would be a good option for the lower skill levels like ''remember'', ''understand'' and ''apply''. For higher skill levels (especially ''create'') preventing access to the internet may not a feasible option, as students typically need access to information for such tasks. Furthermore, another control option is to use \textit{assessment media} that are still \textit{inaccessible} for AI chatbots. For example, asking students to analyse local geodata, interpret maps or statistical figures may be comparably safe, because AI models may not have been trained on such tasks and data\footnote{However, this statement clearly has a limited expiry date, given that multi-modal AI models are being developed that can easily bridge the gap between images, speech and text \cite{gong2023multimodal}.}.   
    \item \textbf{Control and assess the learning process instead of the obtained skills}. For example, closely supervise the way students arrive at a result and rate their learning based on observing this process. This might be an option to rate higher skills levels (''create'', ''evaluate'' and ''analyze'').
    \item \textbf{Forbid the use of chatbots}, check compliance with these rules a-posteriori (there is software providing a likelihood that a text is written by a chatbot\footnote{e.g., \url{https://openai-openai-detector.hf.space/}}), and punish in case of non-compliance. This applies to all levels but may also be the least effective measure; detection software is currently not capable of detecting a large part of the AI-generated text.
\end{enumerate}

Our analysis specifically focused on exams and assignments in Geography teaching at university level; it does not imply that chatbots cannot be beneficial for teaching at all. ChatGPT offers great assistance in e.g. scripting tasks and may replace browsing through help-functions or personal assistance, but only when the right question is asked. A student needs to be able to evaluate the quality of the outcome which requires a thorough understanding of the content and background knowledge. Thus, there is a place for AI-chatbots in education, but including it in examination introduces a major risk for the validity of assessments. 

The discussion around the use of ChatGPT in university teaching adds a new dimension to ethical behaviour in AI, science, and education in general, which is also part of the university learning outcomes. Geography teachers specifically have a responsibility to teach ethics in combination with spatial thinking in the changing global environment, since failing to do so would ''deprive incoming generations
of leaders and bystanders of capabilities essential for navigating uncertain environmental futures'' \cite{larsen2022}. 
Guidelines, such as given in Figure 1 of UNESCO's Quick start guide \cite{unesco2023}, which is a flow diagram asking the reader whether quality of the output matters, they have expertise on the topic, and are willing to take full responsibility of inaccuracies, may help teachers decide which of the actions above is most applicable to the course or assignment at hand, and may at the same time raise awareness with students what they may be sacrificing when using a chatbot. In the end, ''ideas inform ethics, ethics inform policies, and policies make places'' \cite{larsen2022}, and this flow of action requires people to be able to create, evaluate, and analyze independently.









\bibliography{lipics-v2019-sample-article}

\appendix

\end{document}